# Hierarchical Watermarking Framework Based on Analysis of Local Complexity Variations

Majid Mohrekesh[1], Shekoofeh Azizi[2], Shahram Shirani[3], Nader Karimi[1], and Shadrokh Samavi[1,3]
[1]Department of Electrical and Computer Engineering, Isfahan University of Technology, Isfahan, 84156-83111, Iran
[2]Department of Electrical and Computer Engineering, University of British Columbia, Vancouver, V6T 1Z4 Canada
[3]Department of Electrical and Computer Engineering, McMaster University, Hamilton, L8S 4L8 Canada

***Abstract*—Increasing production and exchange of multimedia content has increased the need for better protection of copyright by means of watermarking. Different methods have been proposed to satisfy the tradeoff between imperceptibility and robustness as two important characteristics in watermarking while maintaining proper data-embedding capacity. Many watermarking methods use image independent set of parameters. Different images possess different potentials for robust and transparent hosting of watermark data. To overcome this deficiency, in this paper we have proposed a new hierarchical adaptive watermarking framework. At the higher level of hierarchy, complexity of an image is ranked in comparison with complexities of images of a dataset. For a typical dataset of images, the statistical distribution of block complexities is found. At the lower level of the hierarchy, for a single cover image that is to be watermarked, complexities of blocks can be found. Local complexity variation (LCV) among a block and its neighbors is used to adaptively control the watermark strength factor of each block. Such local complexity analysis creates an adaptive embedding scheme, which results in higher transparency by reducing blockiness effects. This two level hierarchy has enabled our method to take advantage of all image blocks to elevate the embedding capacity while preserving imperceptibility. For testing the effectiveness of the proposed framework, contourlet transform (CT) in conjunction with discrete cosine transform (DCT) is used to embed pseudorandom binary sequences as watermark. Experimental results show that the proposed framework elevates the performance the watermarking routine in terms of both robustness and transparency.**

## I. Introduction

With the ease of access to the internet in recent years, sharing digital media has become easier and faster. Watermarking is an effective way to preserve copyrights and media owner's intellectual property and consequently many watermarking techniques have been proposed. Digital image watermarking techniques embed the owners' copyright information into an image in embedding phase and later on, can extract this information [1]. The efficiency of these techniques depends on perceptual invisibility as well as its robustness against intentional and unintentional attacks. It is essential to note that concurrently satisfying both robustness and invisibility requirements is a challenging problem. Blind watermarking techniques, which are more attractive, do not require the original image for their extraction phase, as opposed to non-blind techniques. Recent techniques use transform domain of images for embedding rather than using the less robust spatial domain [1].

In recent years, several transforms domain image watermarking algorithms have been proposed. Most of these techniques usually employ Discrete Cosine Transform (DCT) [2], Discrete Fourier Transform (DFT) [3], Discrete Wavelet Transform (DWT) [4] and Contourlet Transform (CT) [5]-[8]. Among the transform domain techniques, DCT and DWT based techniques are more popular. DCT based methods are robust against simple image processing attacks and JPEG compression, but unfortunately these methods are not robust to basic transformations such as cropping and resizing [1]. Since DWT has a number of advantages over DFT and DCT, it is widely used in watermarking algorithms. In general, DWT based methods use middle or high frequency regions for embedding of the watermark [4]. Despite these advantages, DWT has some limitations in capturing the directional information, which is addressed by CT [9]. Impressive properties of CT motivated researchers to apply this transform for watermarking purposes. In [5], a non-blind CT based method for image watermarking is proposed. This method embeds the watermark into pixels corresponding to high frequency coefficients of CT and the number of these coefficients is related to the size of the watermark. Moreover, authors of [6] note that CT based methods outperform DWT and DCT based techniques. Song et al. proposed a CT-based image adaptive watermarking scheme in which the watermark is embedded into the largest detail subband of the image [6]. The method presented in [7] is a non-blind method that embeds a watermark in two scales of contourlet transform. In [8], authors present a CT-based watermarking scheme that embeds watermark in directional subband image with highest energy.

In some recent studies a combination of frequency-domain transforms are cascaded to increase robustness of watermarking schemes [10]-[14]. In [10] a DCT-DWT domain method has been proposed. This method is a dual transform-domain watermarking scheme based on the orthogonal components of image sub-spaces which provides a robust authentication process. The method presented in [11] uses singular values of wavelet coefficients and the method in [12] uses values of Hadamard coefficients. There are also methods that embed watermark in DCT coefficients of CT. To elevate robustness, [13] presents a hybrid method that uses DCT coefficients of CT. This hybrid method distributes effects of the embedding by diffusing the changes throughout CT coefficients. In [14] a non-blind watermarking algorithm is proposed for embedding information into medical images. This scheme is another hybrid CT-DCT based method, which embeds the watermark into lowpass subband of an image.

Adaptive image watermarking algorithms specify the location and embedding capacity for watermarking according to the characteristics of the original image, such as complexity, texture, and brightness [15]-[21]. One of the earliest methods of adaptive watermarking is presented in [22], which employs a regional perceptual classifier to assign a noise sensitivity index to each region. This method uses average gradient magnitude for spatial adaptive placement of watermark. After that, in [23] and [24] adaptive watermarking method based on HVS using DCT is presented. Determining watermarking parameters such as the strength factor is another goal of adaptive methods. In [24] a DCT based method using the addition of watermark is proposed. This method classifies blocks of the original image based on visual characteristics of each block. Then, strength factor of embedding is adaptively adjusted for that block. The method presented in [25] is an adaptive blind watermarking algorithm based on image content. This method uses Ridgelet transform to extract where watermark should be embedded and watermark strength factor is adaptively changed based on different image features. In [26], authors present a wavelet based method that adjusts the location of embedding and strength factor according to the characteristics of image. Moreover, authors in [27] employ genetic algorithm to find proper strength factor and control imperceptibility of method.

Almost all non-adaptive watermarking methods propose an embedding scheme for better imperceptibility or higher robustness or a tradeoff of these two. These methods are image independent and each offer a bundle of parameters, which are analytically or empirically obtained. Empirical parameters usually offer no proves for use and analytic alternates often require a huge deal of calculations to reach their final goal. Moreover, wide variations among images demand image-based and dynamic selection of parameters for a desired embedding scheme. Moreover, a watermarking algorithm can adopt two basic types of mechanism for embedding a watermark in the image: (a) the spread spectrum and (b) relationship enforcement based watermarking. The spread spectrum techniques add a noise-like watermark to an image and they detect the watermark via a correlator [28]. On the other hand, most of the methods that belong to the second type use quantization based watermarking [29]-[30].

In this paper, we propose a framework for adaptive watermarking that first tries to find a proper initial embedding parameter and then adaptively change the parameter based on regional characteristics of the image. As a basic watermarking parameter, strength factor ($\alpha$) is used for adaptivity. Higher values of $\alpha$ cause higher robustness. Those methods that use a constant strength factor need to choose a mid-size value for $\alpha$ to obtain a mid-point in the conflicting spectrums of robustness and transparency. Hence, in comparison with constant $\alpha$ methods, our adaptive scheme produces higher robustness in complex areas of the image and higher transparency in smooth areas. A measure of complexity is needed to find out which areas of image are more and which areas are less complex.

Figure 1 shows a diagram of the proposed watermarking framework. First, a set of typical images go into "dataset complexity assessment" box. The proposed complexity assessment measure plays a basic role in this part of the framework. This complexity measure is also used in boxes that are labeled as "single image complexity assessment" and "block complexity assessment". The output of "dataset complexity assessment" box is the statistical data of the image dataset and will be used to rank the desired cover image in "single image complexity assessment" box. For the cover image an initial strength factor, $\alpha_i$, is produced and is fed to "block complexity assessment" box. The cover image and its $\alpha_i$ are sent to the "embedding stage" where for each image block, based on its regional complexity, a strength factor, $\alpha_m$ is chosen and data is embedded. Adaptive selection of strength factor for each block may cause blockiness in smooth areas. Our method uses local complexity variation (LCV) to smoothly change strength factors of neighboring blocks to avoid blockiness. To prove the functionality of the proposed framework, a hybrid CT-DCT embedding scheme is proposed which transforms the image to different CT subbands and embeds watermark bits in DCT coefficients of these. Our experiments show that intelligent modification of $\alpha$ causes a good tradeoff between the perceptual invisibility and robustness.

The remainder of this paper is organized as follows. Section II describes the fundamental objects of the proposed framework and interactions between its different parts. In section III our proposed embedding scheme using CT and DCT is described. Experimental results are presented in section IV. Finally, section V concludes the paper.

## II. Proposed Framework

In this section, we describe our proposed watermarking framework. A review of watermarking methods reveals the importance of parameters, such as embedding strength factor, $\alpha$, on the overall performance of the algorithm. The value of $\alpha$ directly affects both transparency and robustness, which are the two major characteristics of any watermarking method. For a specific watermarking algorithm, the set of values or range of values dedicated to strength factors should depend on the potential capacity of different regions of the image. This also depends on the total number of watermark bits and the intended robustness and transparency. Moreover, large changes of $\alpha$ in neighboring regions create blockiness and should be avoided. Many watermarking schemes resolve the tradeoff between robustness and transparency by empirically finding values of parameters such as $\alpha$. A systematic framework can find these values and hence lower the overhead of finding proper tradeoffs. One approach to achieve this tradeoff is to consider only one of the two constraints of transparency and robustness in the first step. Then in the second step, the other constraint is satisfied. Selecting transparency as the first step, leads us to find more proper image regions for watermarking. The definition of properness depends on the intended embedding scheme. In all methods a slight amount of change is imposed to the watermarked region. Changes should not be salient and should not attract a human observer attention. Also, induced changes should not cause loss of uniformity or quality of the image. Regions with higher fluctuations have higher capacity of embedding and to preserve transparency regions with lower fluctuations should be embedded with lower capacity. Next, to satisfy the robustness constraint, elevation of embedding strength factor in higher capacity regions is considered. Hence, complexity of a region in an image is to be defined. Then the overall complexity of the cover image should be compared with complexities of dataset images to see whether the image is considered as smooth or coarse and hence choose appropriate initial $\alpha$. This comparison is called inter-image adaptivity. Then image regions are compared with each other to decide on how to change the value of $\alpha$ from one region to the other. This is called intra-image adaptivity where each region is embedded with a different suitable strength factor. In the followings, we present three subsections. In subsection II-A the idea of block-complexity is presented. The proposed framework is independent of such complexity definition and other definitions could also be used. Then in subsection II-B we present the first level of the complexity assessment hierarchy which is the assessment of complexity of an image as compared with that of the images of the database. Then in subsection II-C intra-image complexity assessment is detailed. This is done by ranking of complexity of a block as compared with all block-complexities of that image.

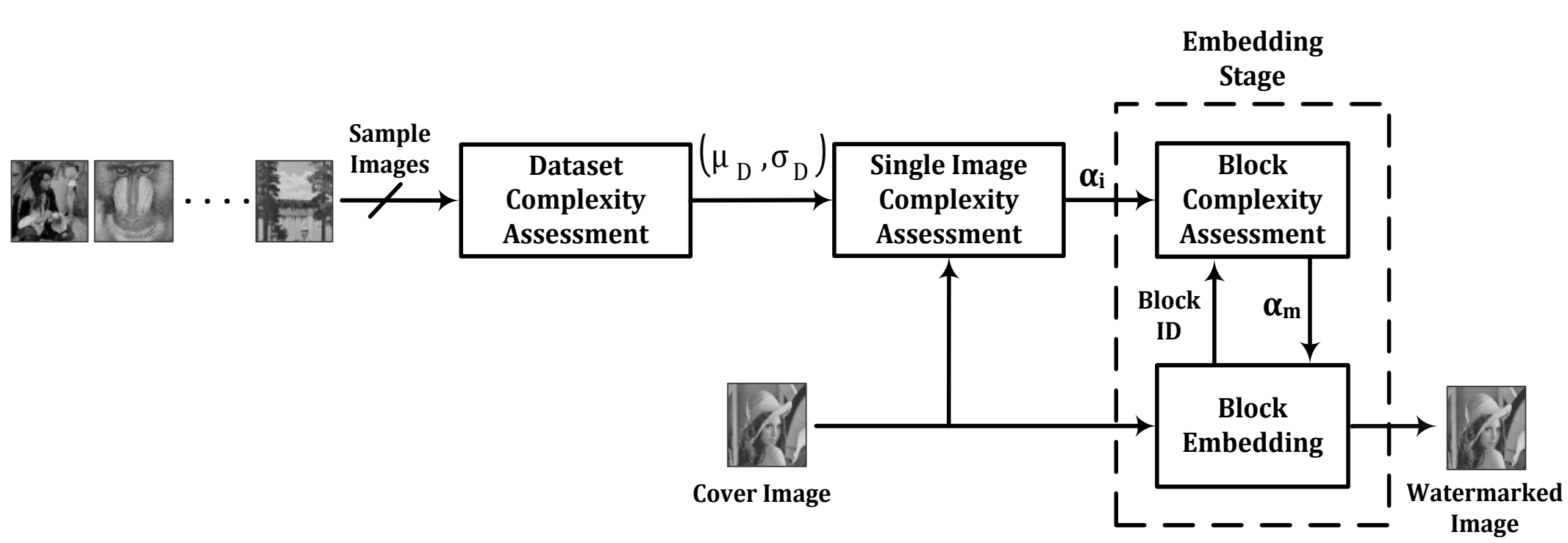

**Fig. 1.** Block diagram of proposed framework.

### A. Complexity Assessment

Regions of image that are more complex could tolerate higher modifications without being noticed by human visual system (HVS). Different means of measuring complexity could be used in spatial or frequency domains. In Equation (1) we present our own definition using a neighborhood of pixels in the spatial domain.

$$C_P[x,y] = \sum_{x'=x-1}^{x+1} \sum_{y'=y-1}^{y+1} \left| l_{x,y} - l_{x',y'} \right| \qquad (1)$$

where $C_P[x,y]$ is the complexity for pixel $P$ at coordinates $[x,y]$ with luminance value of $l_{x,y}$. Also, $l_{x',y'}$ refers to the luminance values of neighboring pixels at coordinates $[x',y']$. Hence, $C_P[x,y]$ calculates the sum of absolute differences of luminance of 8-neighbours of a pixel. A closer look at $C_P$ indicates that it is a kind of texture masking function [28]. Larger values of $C_P[x,y]$ indicate larger fluctuations among neighboring pixels, which means changes in intensity values of such regions would not be noticed by HVS. Our formulation of complexity is not unique. Other formulations can be considered in both spatial and frequency domains as long as they satisfy the relation between complexity and noticeability of change by human visual system. For example entropy has been used as a measure of complexity [20]. Other texture masking functions also could replace the proposed one, but it is relatively simple and has little complexity for calculation, hence its overhead for the watermarking algorithm could be negligible. Figure 2 shows two different image blocks. Both our proposed complexity measure, $C_P$, and entropy, $E$, as measure of complexity are used. Entropy values of both blocks in Fig. 2 are the same while we see that Fig. 2(a) is visually more complex. Our proposed complexity measure correctly gives higher complexity to Fig. 2(a).

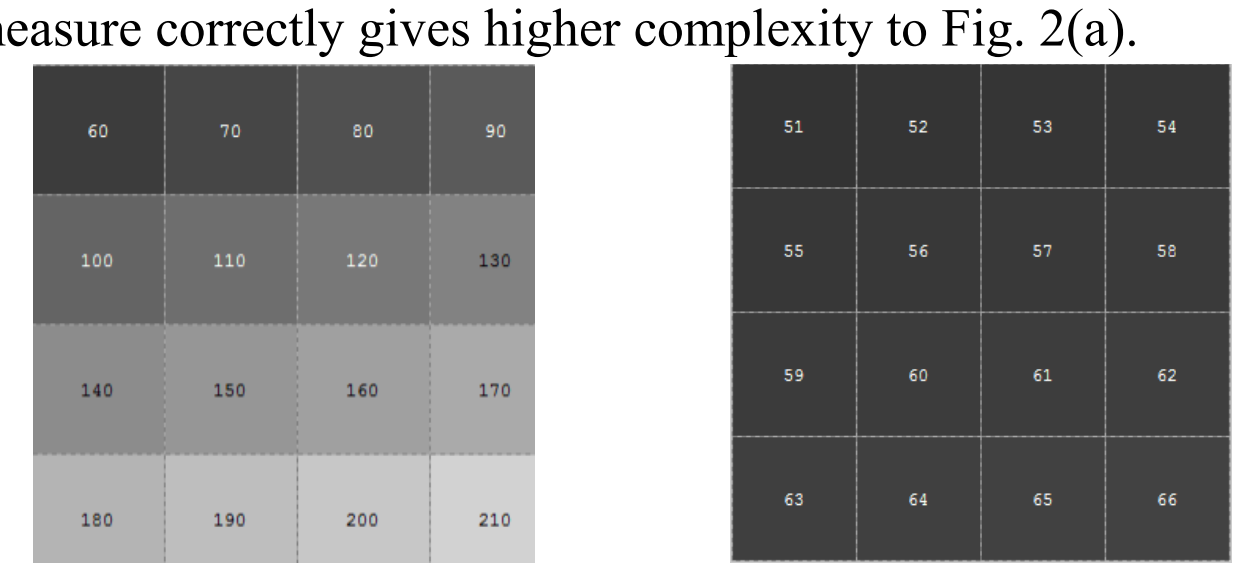

*(a)* $C_P = 260$ , $E = 4$ *(b)* $C_P = 26$ , $E = 4$

**Fig. 2.** Different blocks and operation of complexity assessment measures, $C_P$: proposed complexity measure and E: entropy.

Suppose in every image, $M$ blocks are embedded and blocks are labeled as $block_m$, $m = 1, \ldots, M$. We average $C_P[x,y]$ values of pixels of $block_m$ to get block-complexity, $C(m)$, of that block.

### B. Inter-image Adaptivity

Without loss of generality, consider a watermarking scheme which needs a single strength factor $\alpha$ for its embedding process. Rather than choosing one single $\alpha$ for all images it is better to choose on for each image. If this $\alpha$ were to be picked intelligently for every individual image then it would be expected that better capacity, robustness, and transparency are achieved. Such custom strength factor chosen for image $I_i$ is called $\alpha_i$. We define $\mu_i$, for an image $I_i$, as the average complexity of pixels based on values of $C_P[x,y]$. The procedure for calculating $\alpha_i$ from a dataset of images is shown in pseudo code of TABLE I. In this table $\mu_i$ and $\mu_D$ respectively refer to average pixel complexities in image $I_i$ and in the dataset of images. Also, $\sigma_D$ refers to the standard deviation of pixel complexities in the dataset of images.

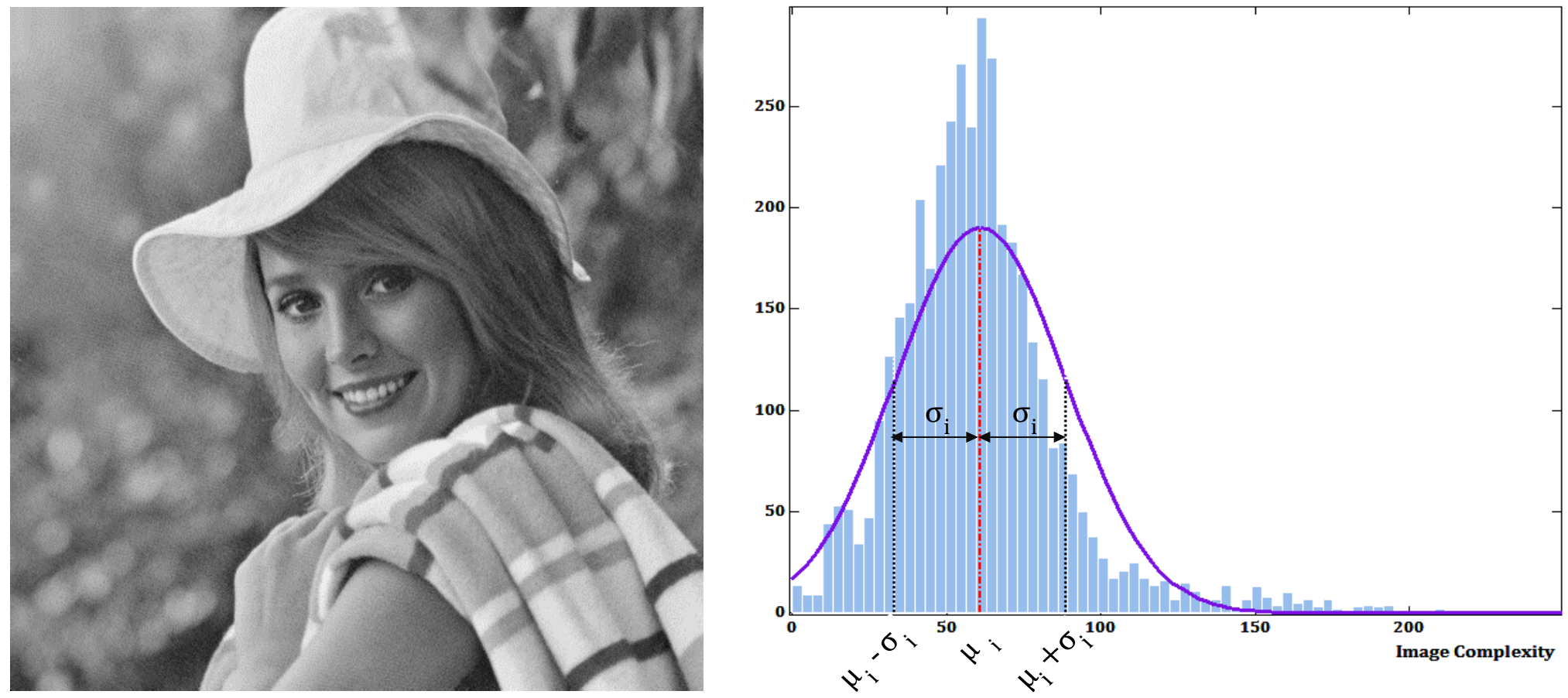

(a) (b)

**Fig. 3.** (a) “Elaine” image, (b) distribution of block-complexities for “Elaine” with a fitted Gaussian envelope.

The strength value $\alpha_0$ is suitable for an image whose $\mu_i$ is close to $\mu_D$ and is in the interval $[\mu_D - \sigma_D, \mu_D + \sigma_D]$. But if an image mean complexity, $\mu_i$, is beyond one $\sigma_D$ from $\mu_D$ then it would be appropriate to change its initial strength factor accordingly. For such images their $\alpha_i$ would be higher or lower than $\alpha_0$, depending on their $\mu_i$. Analysis on “Classic” and “Kodak” datasets show that complexities of most images are in the interval $[\mu_D - \sigma_D, \mu_D + \sigma_D]$. These calculations are shown in block diagram of the framework in Fig. 1 as the “dataset complexity assessment” and “single image complexity assessment”.

TABLE I

PSUEDO CODE TO COMPUTE $\alpha_i$ FOR INTER-IMAGE ADAPTIVTY.

```
Algorithm: Compute α_i
Inputs: i, α_0  Output: α_i
BEGIN
  FOR all I_i in Dataset
    FOR all non-border pixels [x, y] of image I_i
      compute C_P[x, y]
    END
    μ_i ← mean (C_P)
    remove (C_P)
  END
  μ_D ← average (μ)
  σ_D ← standarddeviation(μ);
  α_i ← α_0
  IF |μ_i − μ_D|>σ_D
    α_i ← α_0 × (μ_i/μ_D)
  END
  Return(α_i)
END
```

Robustness is usually evaluated by criteria such as normalized correlation (NC) and bit error rate (BER). In addition, for transparency peak signal to noise ratio (PSNR) has been widely used even though perceptual quality indicators are more suitable. Adapting a different value of $\alpha_i$ for each image causes more robustness for complex images and lower robustness for smooth ones. On the other hand, transparency would be better for smooth images and worse for more complex ones. These tradeoffs cause maintaining mid-range robustness and transparency values for images of a dataset.

### C. *Intra-image Adaptivity*

Suppose that for a given watermarking application both high capacity of embedding and high robustness are needed. Hence, for high robustness a large value of $\alpha_i$ should be used which may cause high distortions in image. We have proposed the intra-image phase to solve this problem. Regions of an image have different complexities. Hence, strength factor should be a function of the complexity of each region. This phase of the framework is done by the “block complexity assessment” box of Fig.1. Hence, for each image an *initial* strength factor, $\alpha_i$, is chosen and used directly for the first block. Thereafter, each block will have an appropriate strength factor, $\alpha_m$, which is derived from local complexity of its neighborhood, but is within an interval around $\alpha_i$. This causes imperceptible embedding even for mostly smooth images. We use blocks of pixels to analyze complexities and to find differences between regions of an image. We traverse the image in a zigzag manner. If $block_m$ were to be embedded, the relative complexity change of $block_m$ as compared with its previous neighbor, $block_{m-1}$, would be calculated. This relative complexity change is used as a criterion for changing of strength factor $\alpha_{m-1}$ and obtaining $\alpha_m$. Figure 3 shows the “Elaine” image with its corresponding distribution of block complexities. Most of blocks are within the interval $[\mu_i - \sigma_i, \mu_i + \sigma_i]$ and would use $\alpha_i$ with no change in its value. The rest of blocks need their custom strength factors ($\alpha_m$).Large changes of $\alpha_m$among neighboring blocks could result in blockiness effects. Hence, blocks are sequentially analyzed and changes in complexity between blocks will be analyzed. This concept is implemented by using relative complexity change factor in Equation (2):

$$\gamma = Relative\ complexity\ change = \frac{C(m) - C(m-1)}{C(m-1)} \quad (2)$$

Figure 4 shows a typical complexity change in a sequence of blocks of the “Pepper” image. Relative changes in complexity values of image blocks cause relative changes in strength factor values of blocks as indicated in Equation (3):

$$Relative\ alpha\ change = \frac{\alpha_m - \alpha_{m-1}}{\alpha_{m-1}} \quad (3)$$

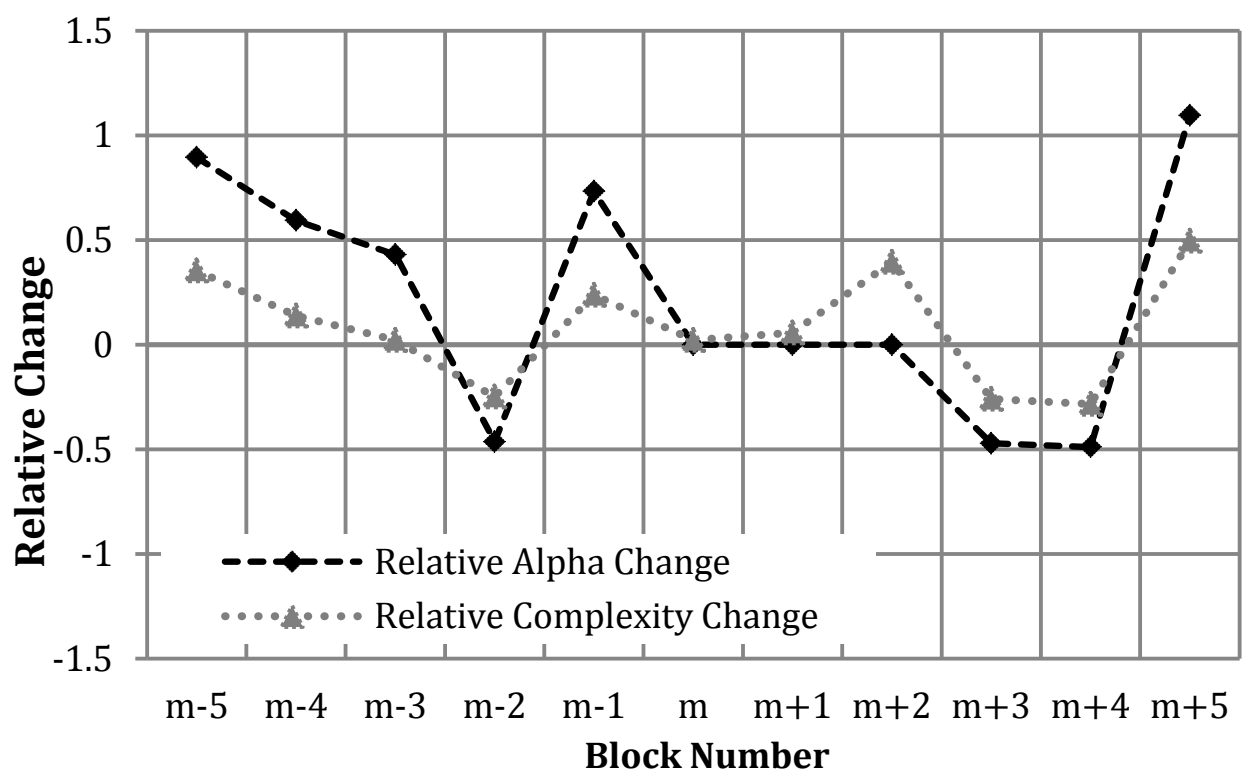

**Fig. 4.** Relative change in complexity values and strength factors in a sequence of blocks in "Peppers" image for $m = 28$. A sequence with increasing complexities results in suppression of strength factors.

Figure 4 demonstrates a situation in which changes in complexities are large which could have resulted in large changes in the magnitude of $\alpha_m$ causing large deviation from $\alpha_i$. Such drastic changes in $\alpha_m$ could have resulted in damaging of either robustness or transparency. Controlled variations in strength values of blocks (changes in $\alpha_m$) and maintaining these values within a reasonable boundary around $\alpha_i$ is obtained by Equation (4):

$$\alpha_m = \begin{cases} \max\ \{(S^{-1}.(1+\gamma).\alpha_{m-1}), (T_1 \times \alpha_i)\} & if\ \gamma < 0 \\ \min\ \{(S.(1+\gamma).\alpha_{m-1}), (T_2 \times \alpha_i)\} & if\ \gamma \geq 0 \end{cases} \quad (4)$$

where $T_2 \times \alpha_i$ and $T_1 \times \alpha_i$ are respectively the upper and lower bounds for the value of $\alpha_m$ and $S$ is a scaling factor. As implied by Equation (2), positive values of $\gamma$ imply growth and negative values imply decrease in complexities. The ratio of complexities of $block_m$ and its previous neighbor $block_{m-1}$ is indicated by $1+\gamma$. This should be also proportional to the ratio of $\alpha_m$ to $\alpha_{m-1}$ in order to force $\alpha_m$ to track the changes of block complexities. A scaling factor $S$, $(S \geq 1)$, forms an interface between two different concepts of complexity and strength factor, and establishes the proportionality between these two concepts. The pair of upper and lower bounds try to contain potential large fluctuations of $\alpha_m$ values within a reasonably small interval.

As an example to visually show the performance of the proposed framework some results are shown in Figure 5. There, we compared a simple embedding method which uses a fixed strength factor of $\alpha = 50$ for all image blocks and we compared it with our adaptive method which has $\alpha_i = 50$. A simple embedding is performed by swapping DCT coefficients of each image block [13]. This swapping technique is explained in details in the next section of the paper. The original "Tiffany" image is shown in Fig. 5(a) and the adaptive embedded image is shown in Fig. 5(b). We see much more artefacts in Fig. 5(c) where the constant strength factor is used. The higher performance of the proposed framework is more apparent when a part of the image is zoomed in. In Fig. 5(e) we have used higher values of $\alpha_m$ in more complex regions while in smooth areas smaller $\alpha_m$ values are used. Hence, in this example, PSNR value for adaptive method is 37.3dB while for the non-adaptive embedding is 35.0dB. Also, we get 0.088 higher structural similarity index (SSIM) value [31]. The average bit error rate (BER) for both methods is less than 4% after JPEG, Salt and Pepper, median filtering and resizing attacks. This means that the watermark was completely retrieved. The values of NC for both methods were more than 0.96 which confirmed the BER results. This shows that the visual quality of our framework is higher while similar robustness is achieved when compared with static strength factor methods. In the following section we will use a more complex embedding method and compare the static version with the proposed adaptive framework.

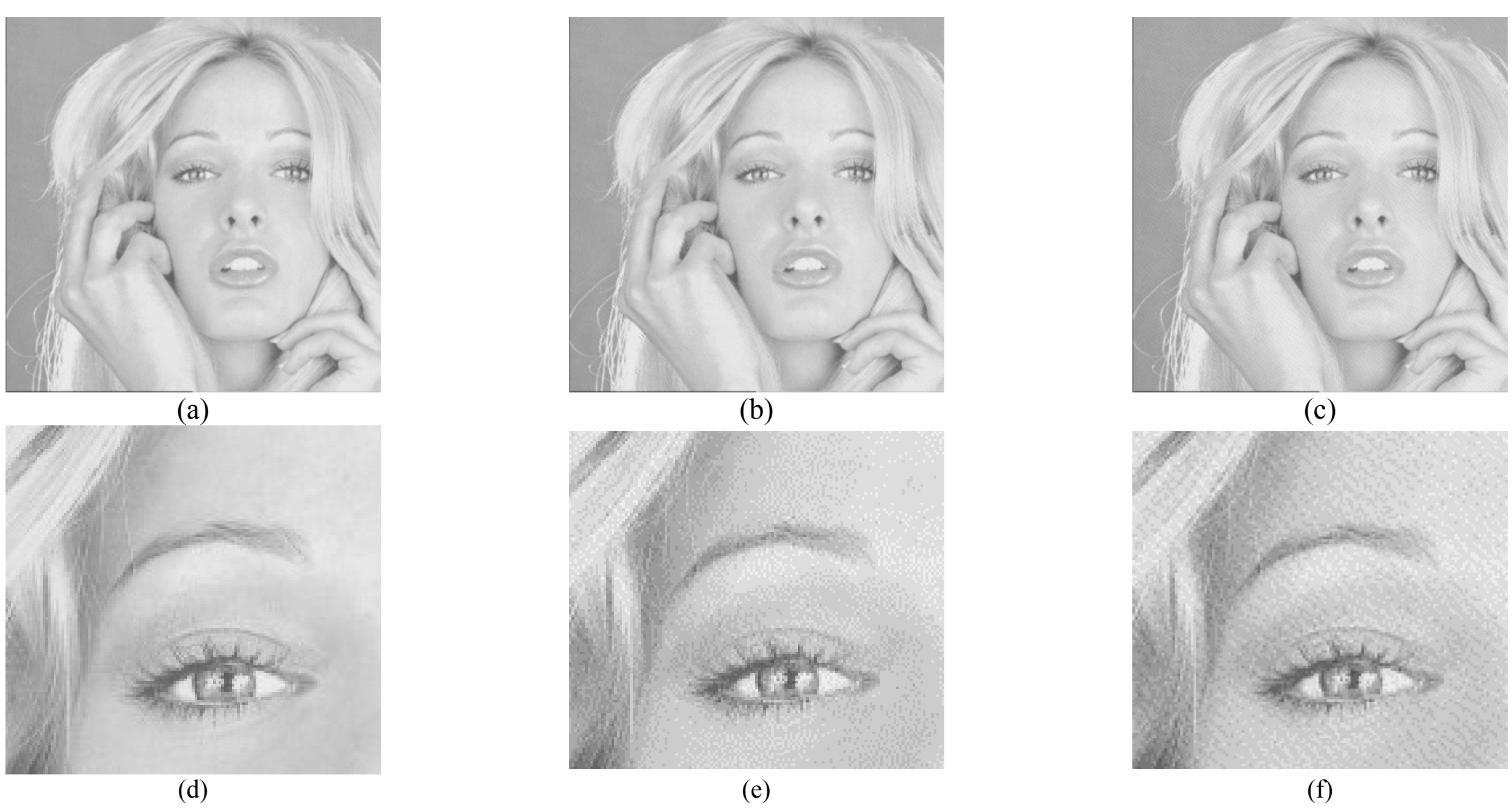

**Fig. 5.** (a) "Tiffany" image, (b) embedded image using our adaptive method, (c) embedded image using non-adaptive method, (d), (e) and (f) zoomed region of corresponding (a), (b) and (c). $\alpha_i$=50 is used for embedding.

The idea of block-based framework is a general one and can be extended to pixel-based methods by considering a block size of 1×1. In such especial case, relative complexity change of a pixel can be calculated and corresponding embedding strength factor can be assigned. A pixel being the smallest element of any image causes the embedding to influence image quality at pixel-level.

So far, we have explained all major steps of our proposed framework. Overall complexity of an image with respect to other images is assessed by inter-image complexity assessment and appropriate $\alpha$ is assigned to the image. In addition, relative complexity variations of blocks of the image are considered by the intra-image adaptivity and by dynamic assignment of $\alpha$ to each bock. In the following section we propose our block embedding method. It should be mentioned that the proposed embedding method does not limit the generality of the proposed framework and other embedding methods could be used instead.

## III. Implementation of Proposed Framework

To implement and test the proposed framework we use proposed complexity measure. Also, for the embedding part of the framework we use an improved version of our previously published scheme [20]. We start by determining a strength factor for each block of the image using the intra-image adaptivity stage. The embedding is done in a cascaded DCT and CT (contourlet transform). The sort of embedding and extraction in this method categorizes it in the group of spread spectrum watermarking methods. This method diffuses the changes among all of the coefficients and results in higher transparency. The strength factor ($\alpha$) varies based on local complexity variations (LCV) of blocks.

Contourlet transform (CT) can efficiently capture smooth contours and edge information of an image in all directions [9]. Do and Vetterli proposed contourlet to overcome deficiencies of previously proposed transforms by new multi-scale and directional representation of images [9]. As shown in Fig. 6, CT consists of two major parts, the Laplacian Pyramid (LP) and Directional Filter Bank (DFB). The LP decomposition at each level generates a low frequency subband image and the difference between the original and the prediction, results in a high frequency (HF) subband image. Subband images from the LP part in different levels are fed into the DFB part where a directional decomposition is performed. Outputs of this part are directional subband images. In one level of Laplacian decomposition of Fig. 6 we see that one low frequency subband image and four directional subband images are produced. Low frequency image is the approximate scale and the four subband images are called detail scale images [9]. Human visual system is less sensitive to minor changes of intensity in complex regions such as edges, thus these image areas are appropriate candidates for watermark embedding. Contourlet transform (CT) represents image edges and provides successive refinements at both spatial and directional resolutions. These characteristics of CT can help identifying image areas where the watermark can easily be hidden with minor distortion. Hence, in this paper we chose CT to exploit these complexity-revealing characteristics.

This watermarking scheme preserves transparency while it is possible to provide more data embedding capacity comparing to comparable methods. All of the blocks of the contourlet space in both approximate and detail scales are used. In the extraction part, the extractor can backtrack the embedding process by just having the secret key. This makes our method a blind watermarking algorithm. The detailed embedding and extraction parts of the algorithm are discussed in the following sub-sections.

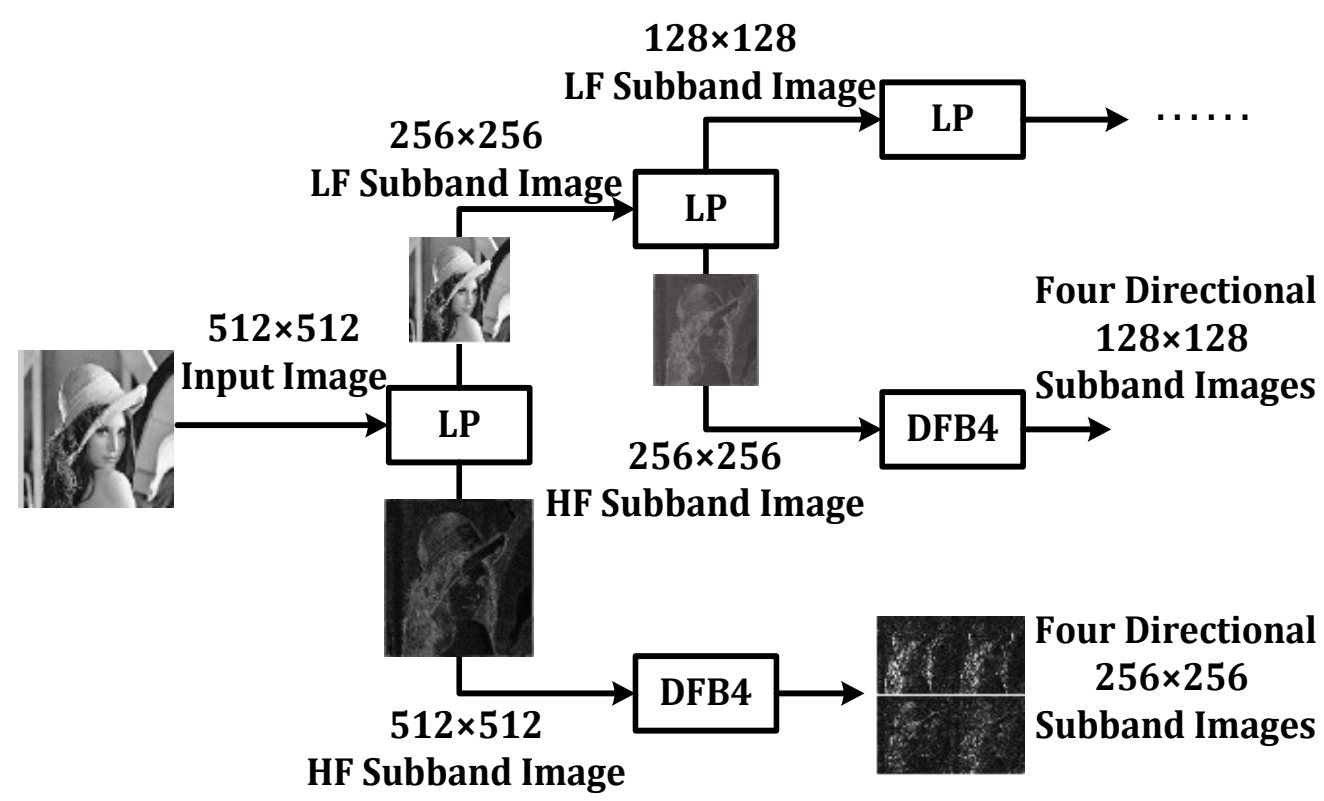

**Fig. 6.** An example of CT consisting of LP and DFB parts in two levels

### A. *Embedding Scheme*

In the embedding part, after transforming the image to contourlet domain, blocks of approximate and detail are fed to the DCT transform module. To satisfy concurrent needs for robustness and transparency, complex blocks are candidate for more robust embedding and smooth ones maintain transparency despite embedding. The watermark is a pseudorandom binary sequence, forming a bit-stream, which is replicated a number of times for redundant embedding to achieve higher robustness. The embedding phase is performed by processing two specific DCT coefficients of the block in a specific order. Then inverse DCT and inverse CT are performed to retrieve the image block and blocks are retiled to form the watermarked image. In this algorithm, inputs include original host image, watermark image, mean strength factor of the dataset ($\alpha_0$) and a secret key. Output is the watermarked image which has good robustness against many attacks and has good transparency. Block diagram of Fig. 7 shows the embedding process. The embedding scheme consists of these major steps:

1. Receive cover image and dataset $\alpha_0$. Calculate the initial strength factor ($\alpha_i$).
2. Decompose original image into approximate and detail scales using CT.
3. Convert watermark into pseudorandom binary sequence and replicate it for redundant embedding in DCT coefficients of contourlet subbands.
4. Partition the CT approximate scale into $L_{AB} \times L_{AB}$ non-overlap blocks and the detail scale images into $L_{DB} \times L_{DB}$ blocks.

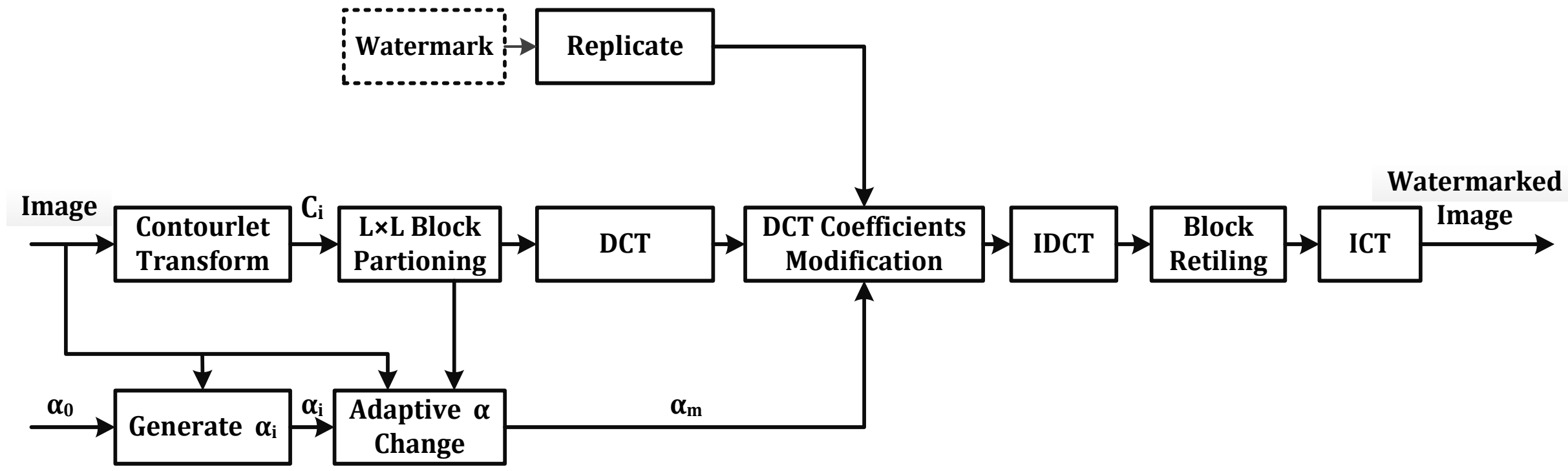

**Fig. 7.** Block diagram of proposed embedding scheme.

5. Scan blocks of each subband in a zigzag manner starting from upper left for odd rows and from upper right for even rows. The zigzag scan order preserves more coherence among subsequent blocks in comparison with raster scan, which has sudden changes in complexities at lines ends. These scanned blocks are fed into the DCT module.
6. Calculate initial strength factor $\alpha_i$ for first block and thereafter, calculate $\alpha_m$ as a function of complexities of previous and current blocks for intra-image adaptivity.
7. Consider two DCT coefficients at positions $(u_A, v_A)$ and $(w_A, z_A)$ of the approximate scale block as candidates for swapping. Also, consider two coefficients at positions $(u_D, v_D)$ and $(w_D, z_D)$ as candidates in the corresponding detail scale images. The order of these coefficients represent a binary bit in the following manner:

$$\begin{cases} if\ \ DCT(u_A, v_A) + \alpha_m < DCT(w_A, z_A) & then\ \ b = 0 \\ if\ \ DCT(u_A, v_A) > \alpha_m + DCT(w_A, z_A) & then\ \ b = 1 \end{cases} \quad (5)$$

   If the current condition of coefficients represent the bit that is to be embedded then no change is required. Otherwise, swap the coefficients and, if needed, add $\alpha_m$ such that their order represents the desired bit from the replicated bit-streams. Strength factor $(\alpha_m)$ is a positive margin, which ensures reliable difference between the two candidate coefficients for robustness against attacks.
8. Return to step 4 if the watermark bit-stream is not finished.
9. Perform inverse DCT and retile blocks to reconstruct approximate and detail scales of contourlet transform.
10. Apply inverse CT for reconstruction of watermarked image.

The replication which mentioned in step 3 causes a redundant embedding of watermark data in the host image to increase the robustness and enhance fidelity of extracted watermark in the presence of attacks. This redundancy is dependent on the algorithm's parameters such as image size $(M \times N)$ and the length of binary sequence $(L_w)$. Suppose the approximate scale of an image in the first level of CT is to be partitioned into $L_{AB} \times L_{AB}$ blocks. Hence, the number of blocks in the approximate scale will be:

$$\#\ of\ Blocks_A = \frac{M \times N}{4 \times {L_{AB}}^2} \quad (6)$$

and the maximum degree of redundancy in this scale would be:

$$Redundancy_A = \frac{\rho}{4 \times {L_{AB}}^2} \quad (7)$$

here $\rho$ is the ratio of the total number of image pixels $(M \times N)$ to the length of binary sequence $(L_w)$. The degree of redundancy in the detail scale is:

$$Redundancy_D = \frac{\rho}{{L_{DB}}^2} \quad (8)$$

where $L_{DB}$ represents the side length of square blocks in the detail scale.

### B. *Extraction Scheme*

The watermarking method that we used is blind and neither the original image nor any side-information is required in the extraction phase. Output of the extraction step is the watermark that was embedded in the original image. Our proposed extraction scheme consists of these steps:

1. Decompose watermarked image to approximate and detail scales using CT.
2. Partition the approximate scale into $L_{AB} \times L_{AB}$ nonoverlap blocks and the detail scale into $L_{DB} \times L_{DB}$ blocks.
3. Scan blocks in a zigzag manner from upper left for odd rows and from upper right for even rows and deliver scanned values sequentially to the DCT module.
4. For each block of scale $s$, watermark bit $b$ is extracted using Equation (9):

$$b = \begin{cases} 0 & if\ \ DCT(u_s, v_s) < DCT(w_s, z_s) \\ 1 & if\ \ DCT(u_s, v_s) > DCT(w_s, z_s) \end{cases} \quad (9)$$

5. After extracting all watermark bits from each scale, pseudorandom binary sequence is reconstructed and the final watermark is obtained by majority weighted voting between intermediate extracted watermarks.

## IV. Experimental Results

For this section, several experiments have been done to evaluate performance of the proposed method. As host images, we used fifteen grayscale classical images including "Lena", "Peppers", "Barbara", "Airplane", "Man", and "Goldhill" with $512 \times 512$ pixels. To extend our experiments we also used 24 grayscale images of Kodak dataset with size of $786 \times 512$ pixels. In addition, 18 images of Canon dataset are tested and a 128-bit payload was used as the embedded data. The results were obtained by averaging over 20 runs with 20 different pseudorandom binary sequences as the watermark. In these experiments, decomposition of image was done with one pyramidal level of CT, which then decomposed into four directional subband images. A pair of "*9-7*" biorthogonal filters was used for both the LP and DFB stages. As mentioned in section III-A approximate and detail scales are partitioned into $L_{AB} \times L_{AB}$ and $L_{DB} \times L_{DB}$ blocks respectively. Each directional subband image has the same size as the approximate scale. In our experiments $\mathrm{L_{AB}}$ is set to 4 and $\mathrm{L_{DB}}$ to 16 to dedicate 4 times relative capacity of embedding to approximate scale. Coordinates of the selected coefficients in each block of approximate scale are $(u,v) = (3,4)$ and $(w,z) = (4,3)$ and the corresponding coordinates in detail scale are $(u,v) = (14,15)$ and $(w,z) = (15,14)$ in cascaded DCT. Starting values of $\alpha_0$ were respectively 11 and 9 in the approximate and detail scales. Parameters to control strength factor $\alpha_m$ were empirically chosen as $S = 1.1, T_1 = 0.5$ and $T_2 = 1.5$. MATLAB 7.12.0 has been the implementation platform.

To evaluate performance of our proposed framework for adaptive watermarking, we compare visual quality and robustness of adaptive and non-adaptive scheme with each other. In non-adaptive scheme strength factor in all of blocks and all of images are constant and is equal to initial strength factor. For fair comparison all parameters in non-adaptive scheme are the same as the proposed adaptive one.

### A. Visual Quality

Figure 8 illustrates original test images and watermarked images using the proposed adaptive method for a message length of 1024 bits. It can be seen that the watermarked images using the proposed approach have high perceptual qualities. In addition, our method has consistent performance both in high and low texture parts of images.

Even for small watermark strings of size 128 bits, our framework shows superior visual quality as compared with non-adaptive comparable method. For example when embedding into the Peppers image, the output watermarked image of our framework has a PSNR of 46.508dB. Using non-adaptive comparable method when the same 128 bits of data was embedded into the Peppers image, we got a PSNR of 45.450dB, which is 1.058dB lower than our method. Moreover, when embedding longer bit-streams, such as 1024 bits of data, we get higher overall PSNR values. For Kodak images, we get an average PSNR of 42.414dB when our framework is used as compared to 41.955dB when non-adaptive comparable embedding is used.

### B. Robustness against Attacks

To evaluate the robustness of the proposed embedding method as a part of the proposed framework, watermarked images were tested against various categories of attacks, such as, geometrical, noising, denoising, compression and image processing attacks. Specific attacks included Rotation (R),

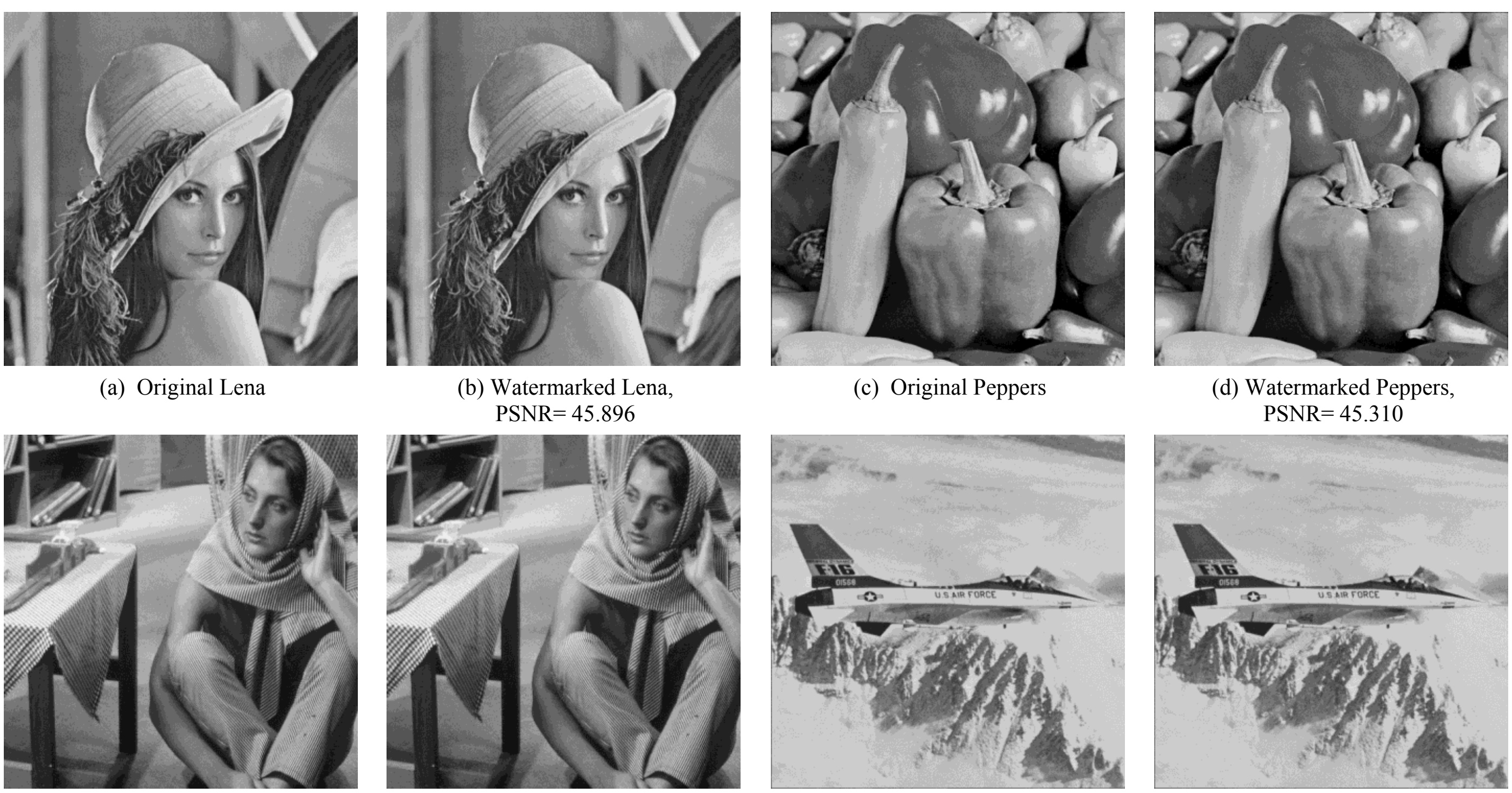

(a) Original Lena
(b) Watermarked Lena, PSNR= 45.896
(c) Original Peppers
(d) Watermarked Peppers, PSNR= 45.310
(e) Original Barbara
(f) Watermarked Barbara, PSNR= 40.550
(g) Original Airplane
(h) WatermarkedAirplane, PSNR= 47.694

**Fig. 8.** (a), (c), (e) and (g): Original images, (b), (d), (f) and (h): corresponding watermarked images using proposed method with 1024 bits message length.

Cropping (C), Resizing (RS), addition of Gaussian Noise (GN), Salt and Pepper noise (S&P), Median Filter (MF), Histogram Equalization (HE), Gamma Correction (GC) and Sharpening (SH). For evaluation of robustness, similarities between the original and extracted watermarks were measured by normalized correlation (NC) and bit error rate (BER).

In the first experiment, the proposed technique was tested against JPEG compression with different quality factors. As seen in Fig. 9 the proposed method was highly robust against JPEG compression with different quality factors down to 40%, and still extractable for 30%. Also in this figure, we can see a comparison between the proposed scheme and the non-adaptive version based on average NC values for 20 different runs of embedding in 15 different classic test images. The comparison verified that the adaptive scheme has better robustness than the non-adaptive version in compression attacks.

In the next experiment, we investigated the robustness against geometrical attacks include cropping, rotation and resizing. We assumed the loss of synchronization due to geometric attacks can be compensated by a synchronization technique, so we concentrate only on the distortion due to these attacks [8][16]. Watermarking algorithms take the image and the payload data as inputs and produce the watermarked image as the output. To have a fair comparison between two algorithms it is enough that same image and same payload data are fed into the two comparing algorithms. Hence, we considered a payload of 128 bits for all of the tested methods. This payload is what other algorithms have used to test and report their results. Then we compared the produced outputs in terms of transparency and robustness. If an algorithm uses redundancy, it is jeopardizing the transparency of its output.

As seen in Fig. 10 our proposed scheme had higher robustness against cropping attack when the cropping ratio was less than or equal 50% of the image. In addition, average NC values for 15 different test images with 20 different messages shows that adaptive scheme has higher robustness against this attack as compared with the non-adaptive version.

In TABLE II we are showing produced results for 4 standard images. In addition, mean values of all results obtained from applying the method to all images in the datasets are reported. TABLE II shows high robustness of our scheme under rotation attacks with different angles. In addition, average NC values of the adaptive method were higher than the non-adaptive version. Moreover, for resizing attacks with scaling factors between 0.5 and 2 our scheme can extract watermark completely and with no error. In general, from these experiments we can say that our adaptive scheme has high robustness against geometrical attacks.

TABLE II
NC VALUES OF EXTRACTED WATERMARK UNDER ROTATION ATTACKS

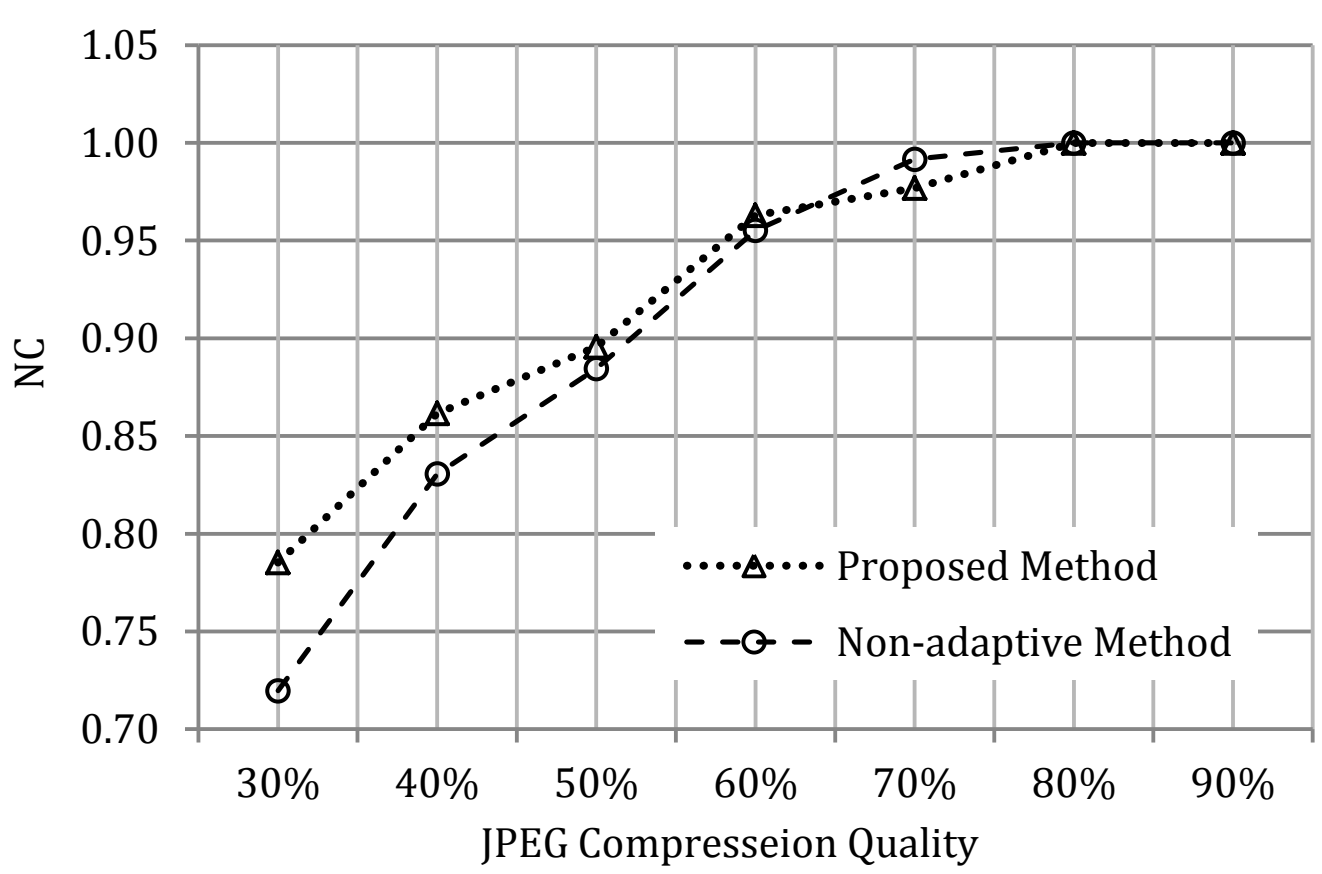

**Fig. 9.** NC values of extracted watermark under JPEG compression attack with different quality factors.

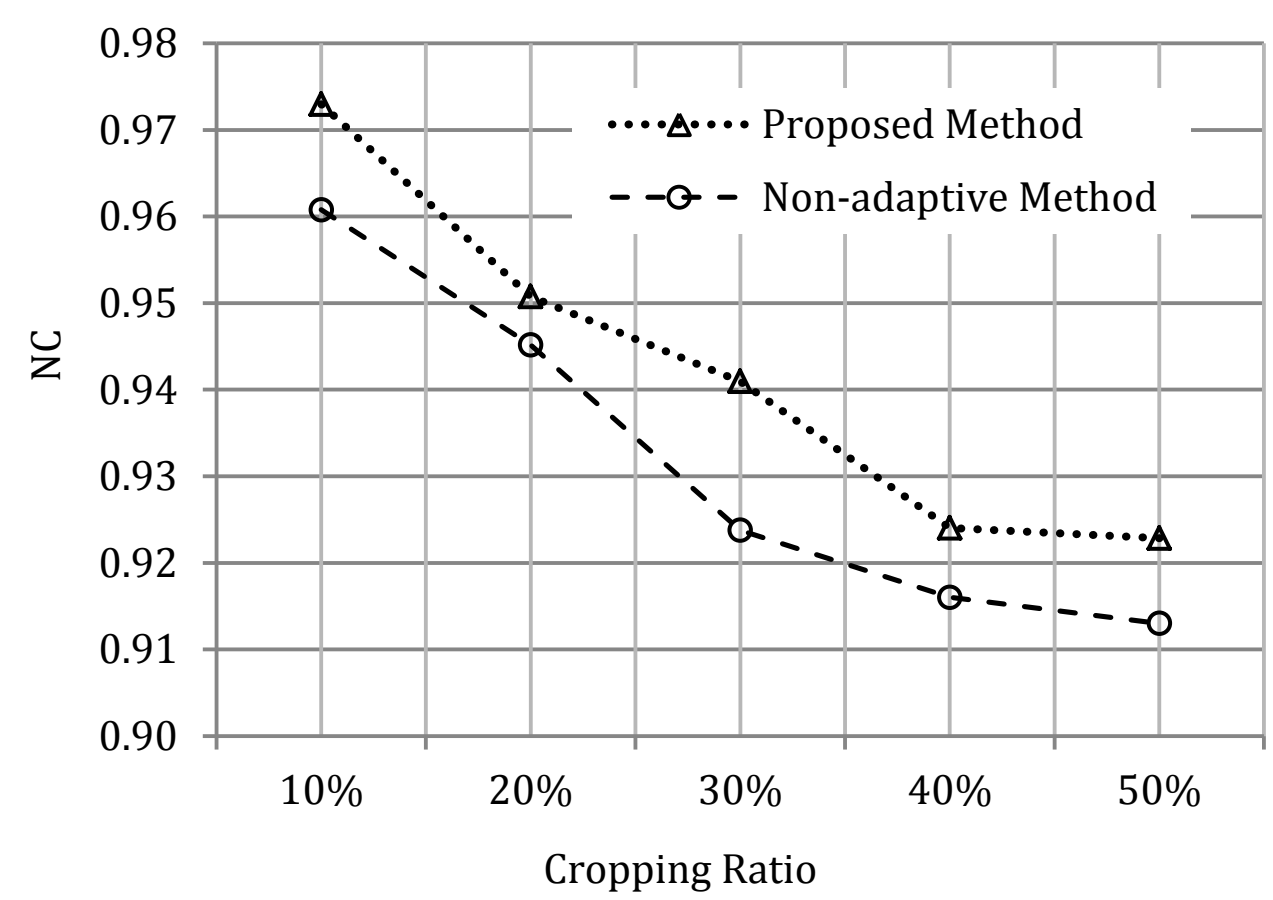

**Fig. 10.** NC of extracted watermark under Cropping attack with different cropping ratios.

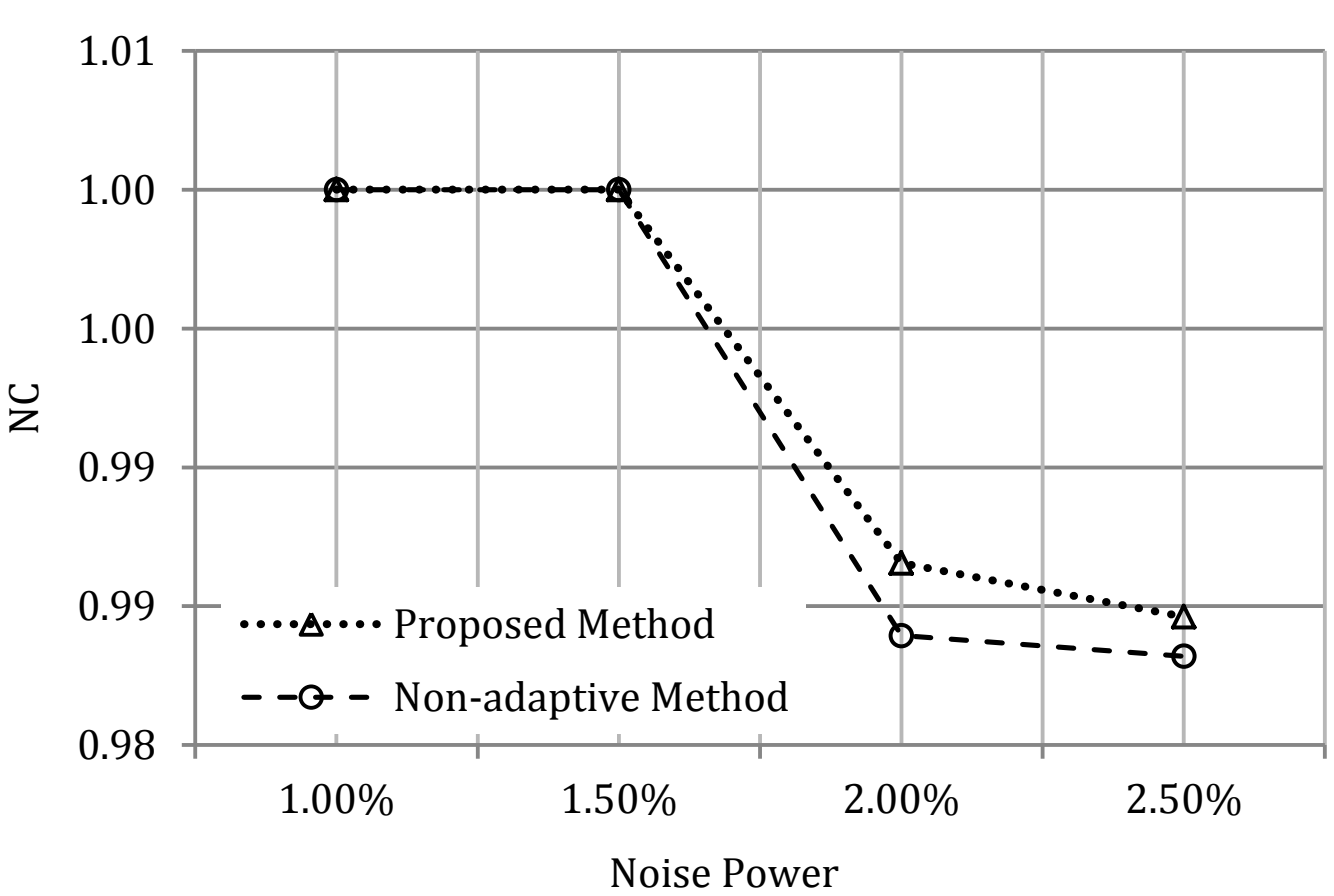

**Fig. 11.** NC values of extracted watermark with added Salt & Pepper noise with different noise densities.

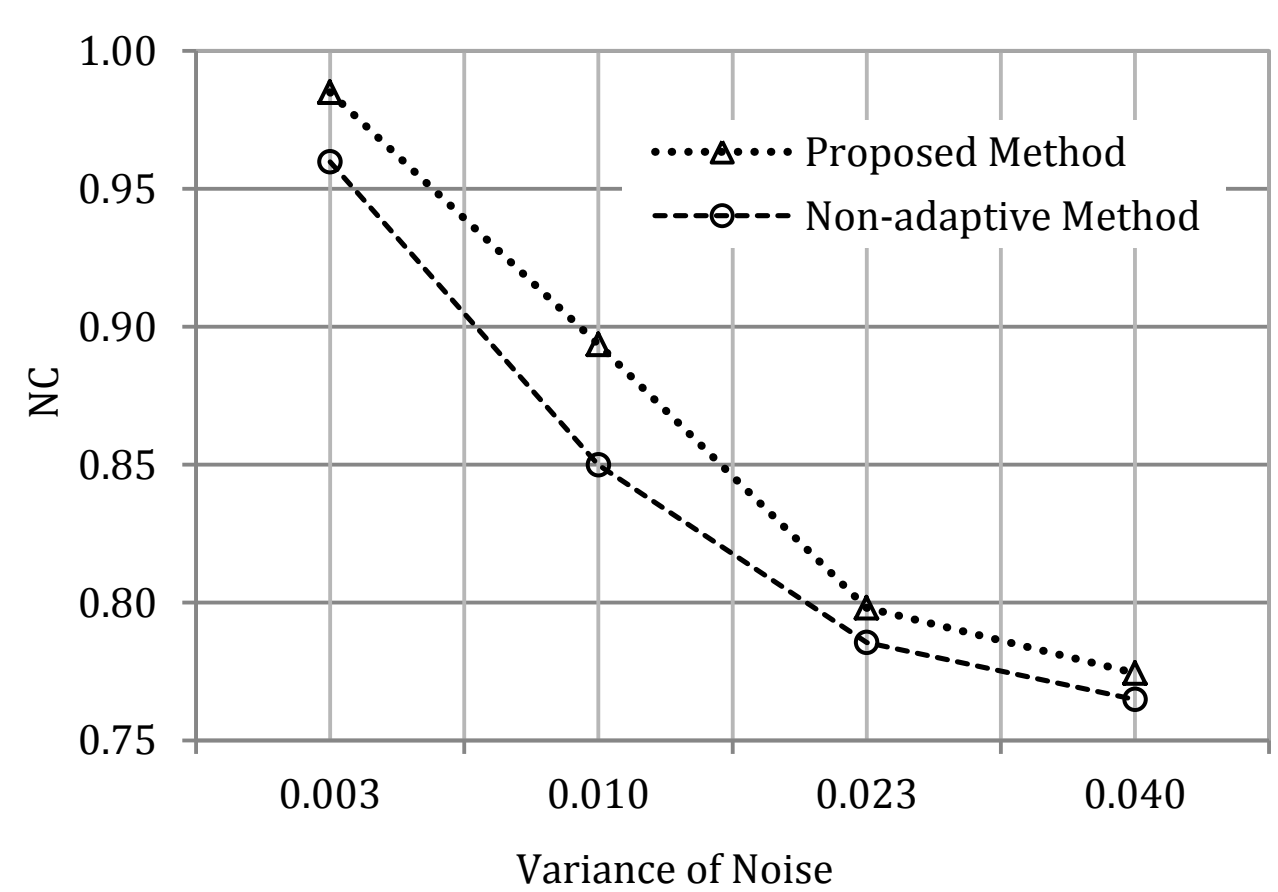

**Fig. 12.** NC values of extracted watermark under additive Gaussian Noise (GN) attack with different noise variances.

(MESSAGE LENGTH = 128 BITS)

| Images (NC) | 0.5º | 1º | 2º | 25º | 45º |
|---|---|---|---|---|---|
| Lena | 0.993 | 0.987 | 0.980 | 0.993 | 1.000 |
| Peppers | 0.953 | 1.000 | 0.980 | 1.000 | 0.980 |
| Barbara | 0.987 | 0.966 | 0.993 | 0.987 | 1.000 |
| Airplane | 0.993 | 1.000 | 1.000 | 1.000 | 1.000 |
| Mean of Adaptive Method | 0.982 | 0.988 | 0.988 | 0.995 | 0.995 |
| Mean of Non-adaptive Method | 0.973 | 0.974 | 0.967 | 0.985 | 0.977 |

TABLE III
NC VALUES OF EXTRACTED WATERMARK UNDER MEDIAN FILTERING ATTACK(MESSAGE LENGTH = 128 BITS)

| Images | Adaptive Proposed Scheme | | | Non-adaptive Scheme | | |
|---|---|---|---|---|---|---|
| | 3×3 | 5×5 | 7×7 | 3×3 | 5×5 | 7×7 |
| Lena | 1.000 | 0.730 | 0.674 | 1.000 | 0.658 | 0.593 |
| Peppers | 1.000 | 0.723 | 0.670 | 1.000 | 0.663 | 0.588 |
| Barbara | 1.000 | 0.780 | 0.671 | 1.000 | 0.660 | 0.588 |
| Airplane | 1.000 | 0.772 | 0.674 | 1.000 | 0.722 | 0.593 |

In the third experiment, we investigated the effect of noising attacks to the proposed adaptive watermarking scheme. For this purpose, we considered additive white GN and Salt & Pepper noise with different noise variances and noise densities. In Fig. 11 and Fig. 12, average NC values of the proposed method and its non-adaptive version are compared for various images under these attacks. The last attack we studied was median filtering attack. TABLE III shows NC results for median filtering with different window sizes for some classic test images. It can be seen that the proposed scheme is highly robust against filtering attacks and compare to non-adaptive version has higher robustness.

TABLE IV and TABLE V shows the resulted NC and BER values for extracted logo after the watermark images were attacked by geometrical, noising, denoising and image processing attacks. Attacks included Rotation 20º and 45º(R), Cropping 10% and 25% (C), Resizing ½ (RS), addition of Gaussian Noise 0.005 (GN), Salt and Pepper noise 0.01 (S&P), JPEG Compression 70% (JC), Median Filter 3×3 (MF), Histogram Equalization (HE), Gamma Correction (GC) and Sharpening (SH). We see that our extracted watermark from the Kodak and Canon dataset have higher average NC and lower average BER values.

### C. Comparison with Other Schemes

To evaluate our method we compared its perceptual quality and robustness with three transform domain state-of-the-art algorithms presented in [8], [15], and [16]. The method of [8] is a recent algorithm which uses CT for embedding purposes and could be considered comparable with our method. The method in [16] is another adaptive method that uses wavelet and has high performance. Also the method in [16] was selected on the basis of its similarity to our method and this method is the nearest competitor to our method. The use of two-layer complexity assessment of the framework is our advantage over [16]. TABLE VI compares PSNR values of watermarked images using proposed adaptive method with its non-adaptive version and those of [8] and [16]. In this table, we are showing results from four images that are used in [8] and [16] as well as the average PSNR that these references have reported. Also, in TABLE VI we are reporting the average PSNR that we have obtained from all images of the dataset. PSNR values of our method are higher than non-adaptive method and method of [8]. Also, in this table average PSNR values of our adaptive method is comparable with method presented in [16]. Although in Goldhill and Airplane we have better PSNR, but our PSNR for Barbara is less that of [16]. These results show that we have higher or comparable perceptual qualities.

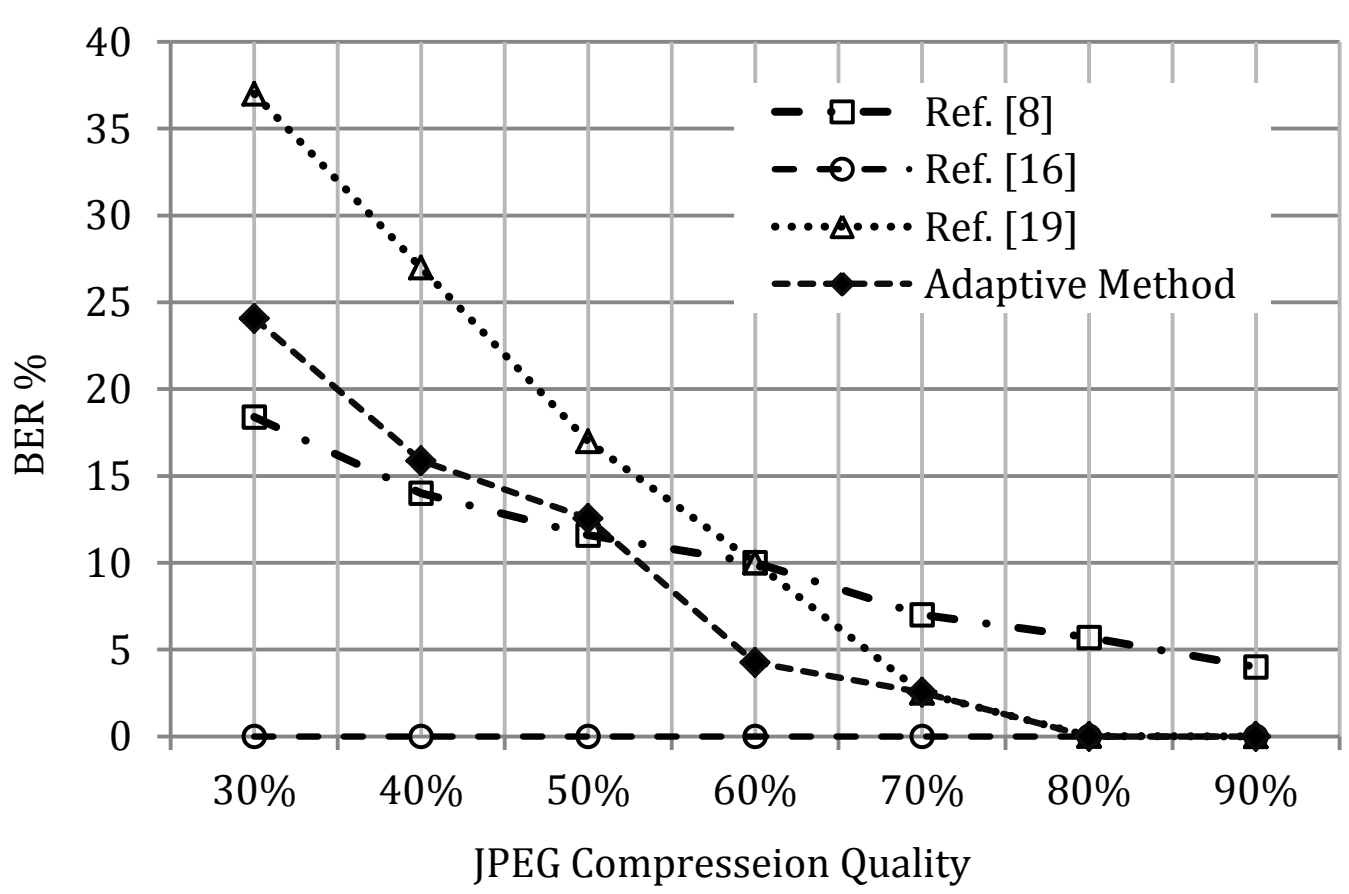

**Fig. 13.** Robustness measure using BER% for comparison of proposed method with [8], [16], and [19] under JPEG attack.

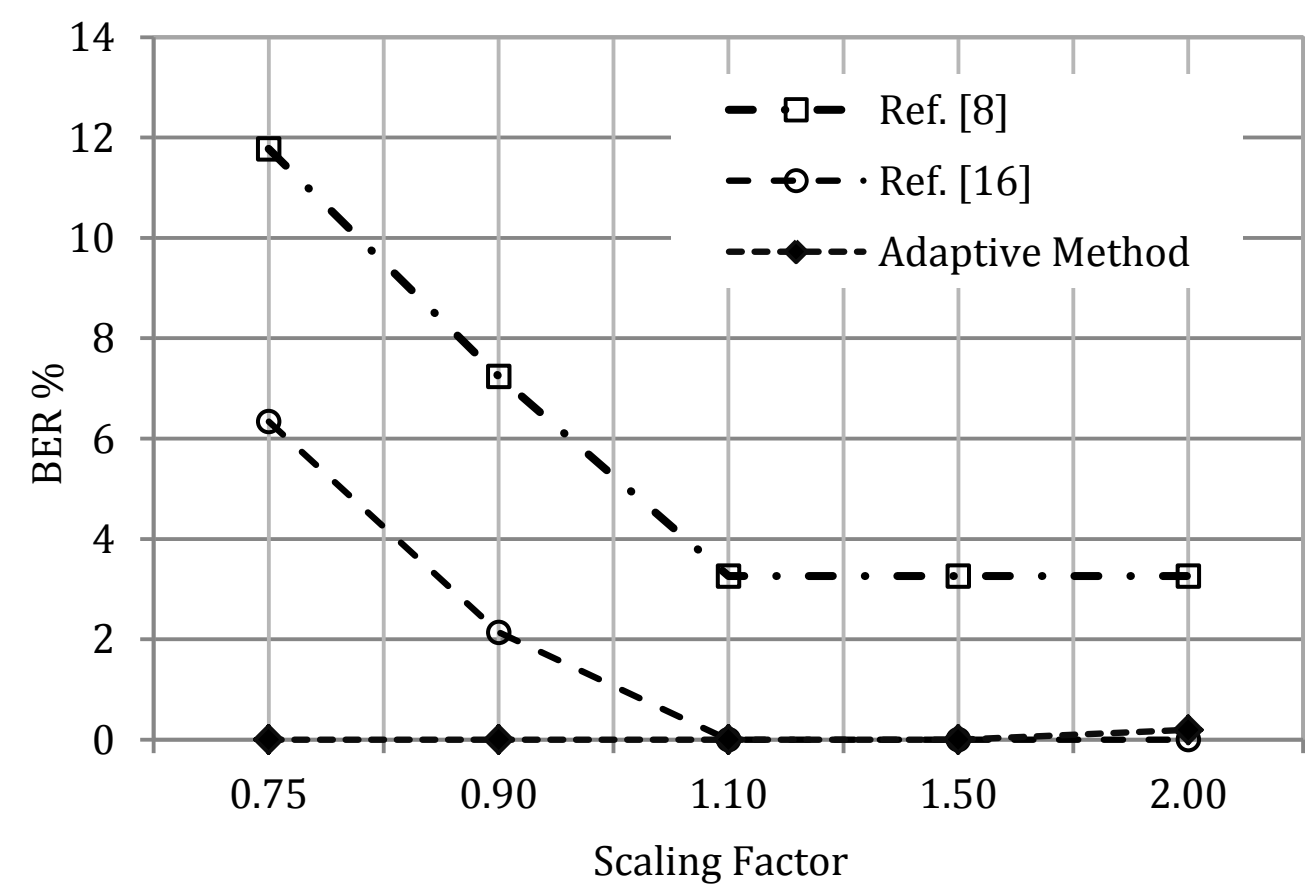

**Fig. 14.** Robustness measure using BER% for comparison of proposed method with [8] and [16] under resizing attack.

TABLE IV
AVERAGE NC AND BER VALUES OF EXTRACTED LOGOS FROM KODAK DATASET USING ADAPTIVE AND NON-ADAPTIVE METHOD
(MESSAGE LENGTH = 128 BITS)

| Methods | Mean | GC | HE | MF | JC | S&P | RS | SH | GN | C10% | C25% | R20º | R45º |
|---|---|---|---|---|---|---|---|---|---|---|---|---|---|
| Adaptive | NC | 1.000 | 0.996 | 0.975 | 0.924 | 0.964 | 0.965 | 0.998 | 0.871 | 0.945 | 0.842 | 0.915 | 0.922 |
| | BER | 0.000 | 0.006 | 0.035 | 0.104 | 0.051 | 0.050 | 0.003 | 0.186 | 0.075 | 0.160 | 0.096 | 0.091 |
| Non-adaptive | NC | 1.000 | 0.992 | 0.949 | 0.922 | 0.912 | 0.961 | 0.997 | 0.847 | 0.750 | 0.628 | 0.845 | 0.875 |
| | BER | 0.001 | 0.011 | 0.069 | 0.105 | 0.118 | 0.053 | 0.004 | 0.202 | 0.303 | 0.419 | 0.203 | 0.188 |

TABLE V
AVERAGE NC AND BER VALUES OF EXTRACTED LOGOS FROM CANON DATASET USING ADAPTIVE AND NON-ADAPTIVE METHOD
(MESSAGE LENGTH = 128 BITS)

| Methods | Mean | GC | HE | MF | JC | S&P | RS | SH | GN | C10% | C25% | R20º | R45º |
|---|---|---|---|---|---|---|---|---|---|---|---|---|---|
| Adaptive | NC | 1.000 | 1.000 | 1.000 | 1.000 | 0.921 | 1.000 | 1.000 | 0.916 | 0.899 | 1.000 | 0.997 | 1.000 |
| | BER | 0.000 | 0.000 | 0.000 | 0.000 | 0.099 | 0.000 | 0.000 | 0.086 | 0.102 | 0.000 | 0.003 | 0.000 |
| Non-adaptive | NC | 1.000 | 1.000 | 1.000 | 0.998 | 0.857 | 0.991 | 0.999 | 0.915 | 0.898 | 0.990 | 0.982 | 1.000 |
| | BER | 0.000 | 0.000 | 0.000 | 0.000 | 0.181 | 0.010 | 0.000 | 0.087 | 0.103 | 0.011 | 0.020 | 0.343 |

TABLE VI
COMPARISON BETWEEN IMPERCEPTIBILITY OF OUR WM FRAMEWORK AND METHODS IN [8]AND [16]: PSNR (dB)

| Method | Goldhill | Barbara | Peppers | Airplane | Average |
|---|---|---|---|---|---|
| Proposed adaptive method | 48.47 | 40.15 | 46.51 | 47.40 | 46.922 |
| Non-adaptive version | 47.90 | 39.58 | 45.45 | 46.46 | 46.216 |
| Ref. [8] | --- | 36.63 | --- | --- | 40.182 |
| Ref. [16] | 47.26 | 48.15 | 47.08 | 45.61 | 46.982 |

TABLE VII
BER (%) COMPARISON OF OUR METHOD WITH [15] AND [16] FOR MEDIAN FILTERING ATTACK (MESSAGE LENGTH = 128 BITS)

| Method | 3×3 | 5×5 | 7×7 |
|---|---|---|---|
| Adaptive method | 00.0 | 00.0 | 42.0 |
| Non-adaptive method | 00.0 | 34.6 | 59.9 |
| Ref. [15] | 52.4 | 54.9 | 60.2 |
| Ref. [16] | 3.20 | 17.9 | ---- |

TABLE VIII
BER (%) COMPARISON OF OUR METHOD WITH [8] AND [16] FOR ROTATION ATTACK (MESSAGE LENGTH = 128 BITS)

| Method | 0.5º | 1º | 2º |
|---|---|---|---|
| Adaptive method | 2.148 | 1.367 | 1.367 |
| Non-adaptive method | 2.930 | 2.734 | 3.516 |
| Ref. [8] | 1.478 | 2.032 | 1.984 |
| Ref. [16] | 1.440 | 1.570 | 1.955 |

As shown in Fig. 13 our method has higher robustness in high quality JPEG compression than [8]. Method of [16] is specifically designed for JPEG compression attacks and has high robustness against such attacks. Method of [16] can extract watermark with no error after compression quality factors of as low as 20%. Also our method can extract watermark with no error after compression with quality factors as low as 70% and for quality factors between 40% and 70% has acceptable BER. As shown in Fig. 14, unlike compression, our adaptive method under resizing attack has better robustness compared to [16] and can completely extract watermark under resizing attack with scaling factors less than 2. These results are achieved due to embedding redundancy. Adaptive change of the strength factor has the potential of embedding with minimal loss of visual quality and higher robustness. Also, employing redundancies in the transform domain further enhances the robustness.

TABLE VII and TABLE VIII compare robustness of our method with selected methods in terms of BER% under filtering and rotation attacks. As seen in TABLE VII for all window sizes, our method has lower BER and higher robustness as compared to [15] and [16]. For most rotation attacks, proposed adaptive method has better performance compared to [8] and [16].

# V. CONCLUSION

In this paper, we proposed a framework for adaptive watermarking to enhance both robustness and imperceptibility of embedding schemes. To prove the functionality of the framework it was implemented and a criterion for measuring local pixel complexity was proposed. The proposed framework is not constrained by the proposed complexity criterion and other means of complexity measure could be applied too. Watermarking adaptivity was achieved by controlling block-based embedding strength factor using block-complexity analysis. The notion of complexity was defined as a relative concept and was considered in a two level hierarchical structure. At the first level of the hierarchy, the general complexity of an image with respect to a large set of standard images is considered. At the second level, complexity of a block in the target image is determined. Strength factor is used as a major controlling parameter in most watermarking schemes. The proposed framework could host any embedding scheme that uses strength factor. This shows the versatility of the framework. To demonstrate that the framework could elevate both robustness and imperceptibility, we improved a hybrid CT-DCT blind embedding method. Comparison between the proposed adaptive and non-adaptive methods showed that the proposed framework was capable of full exploitation of image embedding capacity while keeping high robustness and imperceptibility. We verified higher performance of our method by using PSNR to show imperceptibility, as well as using NC and BER to measure robustness.